\documentclass[letter]{aa}  

\newcommand{\ihor}{$\iota$\,Hor}
\usepackage{graphicx}
\usepackage{txfonts}
\begin{document}

   \title{iota Hor, the first coronal activity cycle in a young
     solar-like star}

   \author{J. Sanz-Forcada\inst{1}
          \and
          B. Stelzer\inst{2} \and T. S. Metcalfe\inst{3,4}
          }

   \institute{Departamento de Astrof\'{i}sica,
     Centro de Astrobiolog\'{i}a (CSIC-INTA), ESAC Campus, P.O. Box 78, 
     E-28691 Villanueva de la Ca\~nada, Madrid, Spain; \\
     \email{jsanz@cab.inta-csic.es}
     \and
     INAF -- Osservatorio Astronomico di Palermo
     G. S. Vaiana, Piazza del Parlamento 1, Palermo, I-90134 Italy
     \email{stelzer@oapa.inaf.it}
     \and
     Space Science Institute, 4750 Walnut Street, Suite 205, Boulder,
     CO 80301, USA
     \email{travis@spacescience.org}
     \and
     Stellar Astrophysics Centre (SAC), Department of Physics and
     Astronomy, Aarhus University, Ny Munkegade 120, DK-8000 Aarhus C,
     Denmark}
   \date{Received 1 March 2013; accepted 8 April 2013 }

 
  \abstract
   {The shortest chromospheric (\ion{Ca}{\sc II} H\&K) activity cycle
     (1.6 yr) has been recently 
     discovered in the young ($\sim600$~Myr) solar-like star
     \ihor. Coronal X-ray activity cycles have only been discovered in a few
     stars other than the Sun, all of them 
     with an older age and a lower activity level than \ihor.}
   {We intended to find the X-ray coronal counterpart of the chromospheric cycle
     for \ihor. This represents the first X-ray cycle observed in an
     active star, as well as the paradigm of the first coronal cycles
     in the life of a solar-like star.}
   {We monitored \ihor\ with XMM-Newton observations spanning almost
     two years. The spectra of each observation are fit with two-temperature
     coronal models to study the long-term variability of the star.}
   {We find a cyclic behavior in X-rays very similar to the
     contemporaneous chromospheric cycle. The continuous chromospheric
     monitoring for
     more than three cycle lengths shows a trend toward decreasing
     amplitude, apparently modulated by a longer term
     trend. The second cycle is disrupted prior to reaching its
     maximum, followed by a brief episode of chaotic variability
     before the cyclic behavior resumes, only to be disrupted again
     after slightly more than one cycle.}
   {We confirm the presence of an activity cycle of $\sim$1.6~yr in
     \ihor\ both in X-rays and \ion{Ca}{\sc II} H\&K. It is likely subject
     to the modulation of a longer, not yet constrained second
     cycle. The 1.6~yr cycle is the shortest coronal one observed to
     date, and \ihor\ represents 
     the most active star for which a coronal activity cycle has been
     found. This cycle is probably representative of the first
     coronal cycles in the life of a solar-like star, at the age when
     life started on Earth.}

   \keywords{stars: activity -- stars: coronae  -- stars: chromospheres 
    -- (stars:) planetary systems --
     astrobiology -- stars: individual: \ihor
                    }

   \maketitle
%


\section{Introduction}\label{sect:intro}

Activity cycles are commonly found among late-type stars through 
chromospheric indicators, such as the Mount Wilson \ion{Ca}{\sc II} H\&K
S-index. \citet{bal95} find that $\sim$60\% of the
main-sequence stars with spectral types from F to early M show
cycles with periods in the range $2.5-25$\,yr. 
Cycles are most often found among moderately active stars.
Very active stars tend to show irregular chromospheric variability
rather than cycles, while inactive stars are found in `Maunder-minimum like'
stages.
The \object{Sun} is well known to display a \ion{Ca}{ii}\,HK cycle of $11$\,yr 
\citep[e.g.][]{white92}, 
which is also observed in the corona through X-ray emission,
spanning more than one dex \citep{orl01,jud03}.

In recent years X-ray counterparts of the \ion{Ca}{ii}\,HK  
activity cycles have been detected in three binary stars: 
HD 81809 \citep{fav04,fav08}, 61\,Cyg\,A \citep{hem06,rob12}, and 
$\alpha$~Cen~B \citep{rob05,rob12,ayr09,dew10}. 
The two components of 61\,Cyg and $\alpha$\,Cen have separations that
are large enough to be resolved in X-rays 
but an activity cycle has so far only been established for one of the two 
stars in each binary. HD\,81809 is a subarcsecond binary and 
is unresolved in both X-rays and \ion{Ca}{ii}\,HK. 
All these stars have long rotation periods, estimated ages of a few Gyrs 
\citep{Barnes07}, and intermediate- to low-activity levels.  
Seasonal changes have been observed in the X-ray emission level of some 
younger and more active stars, such as EK\,Dra 
\citep{gud04} and AB\,Dor \citep{san07}, but
no cyclic variability has been established for them.
The extension of the solar-stellar analogy towards young solar analogs and 
towards more active stars is very important for our understanding of
the early evolution of planets and life on them
\citep[e.g.][]{gui03}. The study of solar analogs in different
evolutionary stages 
indicates that the Sun at a younger age was a faster rotator with a
higher activity level.
Connections between solar activity and Earth's climate
have been discussed in the literature \citep{fri91}. 
A stronger impact is expected for the past, 
given the higher UV flux arriving in the Earth's atmosphere as a result
of the stronger activity \citep{cno07}. 
We present here the first detection of an X-ray activity cycle in a young 
planet-hosting star, \ihor. 

The young solar-like star \ihor\ (HD 17051, \object{HR 810}), has
the mass and $T_{\rm eff}$ of an F8V star \citep{vauc08},  
at a {\em Hipparcos} distance of $17.24 \pm 0.16$\,pc \citep{vanLeeuwen07}. 
It has been proposed as a member of the Hyades cluster
\citep{mon01,met10}, 
which would imply an age of $\sim$625~Myr \citep{leb01}, 
so roughly consistent with the age 
of $\sim$500~Myr calculated from the X-ray emission by \citet{Sanz11},
$\sim$625~Myr from asteroseismology \citep{vauc08},
and 740~Myr from gyrochronology \citep{Barnes07}.
It hosts a planet with 
$M \sin i=2.26\,M_{\rm J}$ orbiting at a separation of $0.92$\,AU 
\citep{Kuerster2000}.
A chromospheric cycle with a period of only $1.6$\,yr has been identified 
by \cite{met10} in \ion{Ca}{ii}\,HK measurements from 
low-resolution spectra collected between 2008 and 2010. This is 
the shortest activity cycle known to date.

Following the discovery of the chromospheric cycle of \ihor,
we started monitoring this star with {\em XMM-Newton} in 2011. The
\ion{Ca}{ii}\,HK observations  
have continued throughout this period. We present here the
  analysis of the long-term variability from the X-ray and
\ion{Ca}{ii}\,HK data collected over the past two years. 
A deeper analysis of the X-ray data will follow in a future publication.
In Sect.\ref{sect:observations} we describe the observations. The results are
given in Sect.~\ref{sect:results}, and Sect.~\ref{sect:disc_and_concl} presents 
the discussion and conclusions. 

%
\begin{table}
\caption[]{XMM-Newton observation log, average exposure times 
  of the three EPIC instruments, and results of spectral
  fits}\label{tab:xraylog}  
\tabcolsep 2.4 pt
\renewcommand{\arraystretch}{1.3}
\begin{center}
\begin{small}
\vspace{-6mm}
\begin{tabular}{lcccc}
\hline \hline
{Date} & {t$_{\rm exp}$} & $\log T$ & $\log EM$ & $L_{\rm X}$\tablefootmark{a} \\
& {(ks)} & (K) & (cm$^{-3}$) & \\
\hline
2011-05-16  &	6.5 & $6.47^{+0.06}_{-0.04}$, $6.82^{+0.03}_{-0.02}$ & $51.12^{+0.08}_{-0.05}$, $50.89^{+0.07}_{-0.17}$ & 6.0 \\
2011-06-11  &  11.5 & $6.54^{+0.01}_{-0.01}$, $6.93^{+0.02}_{-0.02}$ & $51.32^{+0.02}_{-0.02}$, $50.74^{+0.06}_{-0.07}$ & 7.4 \\
2011-07-09  &	8.4 & $6.50^{+0.03}_{-0.03}$, $6.84^{+0.03}_{-0.02}$ & $51.18^{+0.05}_{-0.04}$, $50.84^{+0.07}_{-0.11}$ & 6.3 \\
2011-08-04  &	7.7 & $6.56^{+0.01}_{-0.01}$, $6.96^{+0.03}_{-0.02}$ & $51.33^{+0.03}_{-0.03}$, $50.78^{+0.06}_{-0.07}$ & 7.8 \\
2011-11-20  &	6.9 & $6.56^{+0.02}_{-0.02}$, $6.97^{+0.03}_{-0.02}$ & $51.27^{+0.03}_{-0.03}$, $50.85^{+0.06}_{-0.06}$ & 7.5 \\
2011-12-18  &	6.3 & $6.54^{+0.02}_{-0.02}$, $6.96^{+0.04}_{-0.03}$ & $51.23^{+0.03}_{-0.03}$, $50.60^{+0.08}_{-0.11}$ & 5.8 \\
2012-01-15  &	6.0 & $6.46^{+0.05}_{-0.07}$, $6.82^{+0.02}_{-0.02}$ & $51.00^{+0.06}_{-0.07}$, $50.86^{+0.07}_{-0.10}$ & 5.1 \\
2012-02-10  &	9.0 & $6.47^{+0.03}_{-0.04}$, $6.80^{+0.02}_{-0.03}$ & $51.06^{+0.04}_{-0.05}$, $50.62^{+0.10}_{-0.11}$ & 4.2 \\
2012-05-19  &	7.7 & $6.43^{+0.04}_{-0.04}$, $6.83^{+0.02}_{-0.02}$ & $51.06^{+0.05}_{-0.05}$, $50.73^{+0.07}_{-0.10}$ & 4.6 \\
2012-06-29  &  11.4 & $6.44^{+0.03}_{-0.05}$, $6.83^{+0.02}_{-0.01}$ & $51.09^{+0.04}_{-0.04}$, $50.85^{+0.06}_{-0.06}$ & 5.5 \\
2012-08-09  &	8.1 & $6.44^{+0.04}_{-0.05}$, $6.82^{+0.02}_{-0.01}$ & $51.05^{+0.05}_{-0.05}$, $50.83^{+0.06}_{-0.08}$ & 5.1 \\
2012-11-18  &	5.3 & $6.48^{+0.07}_{-0.05}$, $6.83^{+0.09}_{-0.02}$ & $51.15^{+0.10}_{-0.06}$, $50.78^{+0.09}_{-0.34}$ & 5.6 \\
2012-12-20  &   5.5 & $6.46^{+0.05}_{-0.07}$, $6.83^{+0.02}_{-0.02}$ & $51.04^{+0.06}_{-0.07}$, $50.89^{+0.07}_{-0.09}$ & 5.5 \\
\vspace{0.5mm}
2013-02-03  &   7.6 & $6.46^{+0.04}_{-0.05}$, $6.84^{+0.02}_{-0.02}$ & $51.05^{+0.05}_{-0.06}$, $50.81^{+0.07}_{-0.08}$ & 5.0 \\
\hline
\end{tabular}
\end{small}
\end{center}
\vspace{-6mm}
\tablefoot{\tablefoottext{a}{$L_{\rm X}$ ($\times 10^{28}$ erg\,s$^{-1}$) has
  statistical errors in the range 0.05--0.08.}}
\renewcommand{\arraystretch}{1.}
\end{table}


\section{Observations}\label{sect:observations}

\subsection{Coronal X-rays}\label{subsect:obs_xrays}

Between May 2011 and February 2013 we obtained 14 
snapshots of $\iota$~Hor (XMM prop. ID \#067361, \#069355,
P.I. J. Sanz-Forcada, and a DDT observation on Feb. 3 2013). 
The observing log is presented in Table~\ref{tab:xraylog}. 
Data were reduced following standard procedures in the software SAS
v12.0.1. After removing time intervals affected by high background,
the individual exposure times were between 5\,ks and 11\,ks.
Light curves of the three XMM-Newton/EPIC detectors \citep[spectral range
0.1--15~keV, $E/\Delta E\sim$20--50, ][]{tur01,str01} were extracted
separately for each of the three EPIC 
instruments to identify eventual short-term variations such as flares. 
The ISIS package \citep{isis} and the Astrophysics Plasma Emission Database
\citep[APED v2.0.2,][]{aped} were used to simultaneously fit the
  EPIC (pn, MOS1 and MOS2) spectra of each epoch  
with two-temperature models. Next to the two temperatures, in the
initial fits we let the overall stellar metallicity and the
oxygen abundance be free parameters,
resulting in fits with unconstrained abundances in a few time intervals. 
We then fixed these two parameters
to their median values ([M/H]=0.1, [O/H]=--0.44) of all observations, and
performed a new spectral fit for each observation. The best-fit temperatures
($T_1$, $T_2$) and emission measures ($EM_1$, $EM_2$) are given in
Table~\ref{tab:xraylog}. The spectral analysis also yields the X-ray
flux for the individual observations from which we computed the
X-ray luminosities listed in Table~\ref{tab:xraylog}, calculated in
the canonical ROSAT band 0.12--2.48~keV (5--100~\AA), with error bars 
based on the signal-to-noise ratio. The 
surface flux ($F_{\rm X}$) was calculated using a stellar radius of
$R=1.18$~R$_{\odot}$ \citep{met10}.

\subsection{Chromospheric Ca\,II HK}\label{subsect:obs_calcium}

Observations of the \ion{Ca}{ii} HK lines for $\iota$~Hor were collected as 
part of the SMARTS Southern HK project \citep{met09}, a time-domain survey 
of stellar activity variations for the brightest stars in the southern 
hemisphere. The discovery observations for the 1.6-year activity cycle in 
$\iota$~Hor included 74 spectra spanning 37 epochs between Feb. 15 2008 and 
Aug. 09 2010. Monitoring continued after the discovery with 70 additional 
spectra collected on 35 epochs from Aug. 26 2010 through Feb. 1 2013, 
shortly before the {\it RC Spec} instrument was decommissioned and the 
SMARTS Southern HK project officially ended. Details of the data reduction 
and calibration procedure can be found in \cite{met10}, and the new 
observations are tabulated in the online material (see
Table~\ref{tab:caiilog}). 

\onltab{
%
\begin{table}
\caption[]{\ion{Ca}{ii} H\&K observation log, and measured Mount-Wilson S-index}\label{tab:caiilog}  
\begin{center}
\begin{scriptsize}
\begin{tabular}{lcccc}
\hline \hline
{Date} & {Time (UT)} & BJD\tablefootmark{a} & S-index & $\sigma_S$ \\
& {HH MM SS} & (2450000+) &  & \\
\hline
2010 Aug 26 & 06 24 10 & 5434.76985 & 0.2487 & 0.0028 \\
2010 Aug 26 & 06 25 24 & 5434.77070 & 0.2586 & 0.0033 \\
2010 Sep 16 & 06 08 13 & 5455.75918 & 0.2436 & 0.0044 \\
2010 Sep 16 & 06 09 27 & 5455.76004 & 0.2325 & 0.0040 \\
2010 Sep 21 & 07 33 01 & 5460.81813 & 0.2372 & 0.0025 \\
2010 Sep 21 & 07 34 15 & 5460.81898 & 0.2472 & 0.0028 \\
2010 Oct 15 & 07 00 42 & 5484.79563 & 0.2227 & 0.0032 \\
2010 Oct 15 & 07 01 56 & 5484.79648 & 0.2360 & 0.0041 \\
2010 Oct 20 & 02 42 00 & 5489.61591 & 0.2559 & 0.0019 \\
2010 Oct 20 & 02 43 14 & 5489.61676 & 0.2532 & 0.0019 \\
2010 Nov 09 & 03 53 55 & 5509.66538 & 0.2460 & 0.0017 \\
2010 Nov 09 & 03 55 09 & 5509.66623 & 0.2583 & 0.0019 \\
2010 Nov 26 & 04 05 10 & 5526.67258 & 0.2553 & 0.0021 \\
2010 Nov 26 & 04 06 24 & 5526.67344 & 0.2616 & 0.0022 \\
2010 Dec 13 & 05 12 35 & 5543.71865 & 0.2249 & 0.0018 \\
2010 Dec 13 & 05 13 49 & 5543.71951 & 0.2238 & 0.0018 \\
2010 Dec 31 & 02 47 40 & 5561.61715 & 0.2332 & 0.0026 \\
2010 Dec 31 & 02 48 54 & 5561.61800 & 0.2267 & 0.0026 \\
2011 Jan 08 & 03 02 30 & 5569.62706 & 0.2321 & 0.0019 \\
2011 Jan 08 & 03 03 44 & 5569.62792 & 0.2303 & 0.0018 \\
2011 Jan 30 & 03 34 14 & 5591.64809 & 0.2394 & 0.0018 \\
2011 Jan 30 & 03 35 28 & 5591.64894 & 0.2324 & 0.0017 \\
2011 Feb 10 & 03 07 39 & 5602.62921 & 0.2483 & 0.0021 \\
2011 Feb 10 & 03 08 54 & 5602.63007 & 0.2478 & 0.0021 \\
2011 Feb 26 & 00 57 42 & 5618.53847 & 0.2500 & 0.0030 \\
2011 Feb 26 & 00 58 56 & 5618.53932 & 0.2400 & 0.0028 \\
2011 Mar 09 & 00 49 39 & 5629.53263 & 0.2449 & 0.0020 \\
2011 Mar 09 & 00 50 53 & 5629.53348 & 0.2463 & 0.0022 \\
2011 Apr 04 & 23 21 18 & 5656.47110 & 0.2391 & 0.0019 \\
2011 Apr 04 & 23 22 32 & 5656.47196 & 0.2391 & 0.0018 \\
2011 Jun 10 & 10 25 05 & 5722.93392 & 0.2445 & 0.0020 \\
2011 Jun 10 & 10 26 20 & 5722.93478 & 0.2460 & 0.0021 \\
2011 Jun 29 & 09 59 07 & 5741.91677 & 0.2461 & 0.0024 \\
2011 Jun 29 & 10 00 21 & 5741.91762 & 0.2520 & 0.0023 \\
2011 Jul 20 & 09 09 57 & 5762.88359 & 0.2422 & 0.0025 \\
2011 Jul 20 & 09 11 11 & 5762.88445 & 0.2551 & 0.0026 \\
2011 Sep 01 & 07 59 47 & 5805.83640 & 0.2509 & 0.0024 \\
2011 Sep 01 & 08 01 02 & 5805.83726 & 0.2420 & 0.0022 \\
2011 Sep 30 & 09 38 44 & 5834.90546 & 0.2372 & 0.0022 \\
2011 Sep 30 & 09 39 58 & 5834.90632 & 0.2450 & 0.0022 \\
2011 Oct 22 & 07 21 15 & 5856.80980 & 0.2529 & 0.0036 \\
2011 Oct 22 & 07 22 29 & 5856.81066 & 0.2445 & 0.0033 \\
2011 Nov 14 & 05 51 40 & 5879.74699 & 0.2444 & 0.0018 \\
2011 Nov 14 & 05 52 54 & 5879.74785 & 0.2373 & 0.0017 \\
2011 Dec 12 & 02 22 47 & 5907.60081 & 0.2387 & 0.0024 \\
2011 Dec 12 & 02 24 02 & 5907.60166 & 0.2442 & 0.0024 \\
2011 Dec 23 & 02 15 40 & 5918.59534 & 0.2165 & 0.0052 \\
2011 Dec 23 & 02 16 54 & 5918.59619 & 0.2177 & 0.0043 \\
2012 Jan 14 & 04 19 10 & 5940.68003 & 0.2319 & 0.0017 \\
2012 Jan 14 & 04 20 24 & 5940.68089 & 0.2292 & 0.0017 \\
2012 Jan 28 & 02 27 14 & 5954.60166 & 0.2285 & 0.0016 \\
2012 Jan 28 & 02 28 28 & 5954.60252 & 0.2289 & 0.0016 \\
2012 Feb 14 & 01 39 13 & 5971.56766 & 0.2337 & 0.0030 \\
2012 Feb 14 & 01 40 27 & 5971.56851 & 0.2201 & 0.0030 \\
2012 Jul 09 & 08 38 20 & 6117.86119 & 0.2218 & 0.0015 \\
2012 Jul 09 & 08 39 35 & 6117.86204 & 0.2285 & 0.0015 \\
2012 Aug 09 & 09 13 10 & 6148.88669 & 0.2226 & 0.0017 \\
2012 Aug 09 & 09 14 24 & 6148.88755 & 0.2263 & 0.0016 \\
2012 Sep 03 & 08 53 51 & 6173.87402 & 0.2363 & 0.0024 \\
2012 Sep 03 & 08 55 05 & 6173.87488 & 0.2293 & 0.0026 \\
2012 Oct 13 & 07 07 34 & 6213.80043 & 0.2321 & 0.0020 \\
2012 Oct 13 & 07 08 48 & 6213.80129 & 0.2394 & 0.0020 \\
2012 Nov 30 & 05 29 13 & 6261.73078 & 0.2365 & 0.0055 \\
2012 Nov 30 & 05 30 27 & 6261.73164 & 0.2306 & 0.0070 \\
2012 Dec 11 & 03 13 41 & 6272.63618 & 0.2404 & 0.0028 \\
2012 Dec 11 & 03 14 55 & 6272.63703 & 0.2453 & 0.0027 \\
2012 Dec 30 & 04 53 05 & 6291.70428 & 0.2298 & 0.0033 \\
2012 Dec 30 & 04 54 19 & 6291.70514 & 0.2402 & 0.0034 \\
2013 Feb 01 & 02 20 02 & 6324.59648 & 0.2340 & 0.0026 \\
2013 Feb 01 & 02 21 16 & 6324.59734 & 0.2211 & 0.0019 \\
\hline
\end{tabular}

\end{scriptsize}
\end{center}
\tablefoot{
\tablefoottext{a}{Barycentric Julian Date of observation}}
\end{table}
}

%
\begin{figure}[t]
  \centering
  \vspace{0.5cm}
  \includegraphics[angle=90,width=0.45\textwidth]{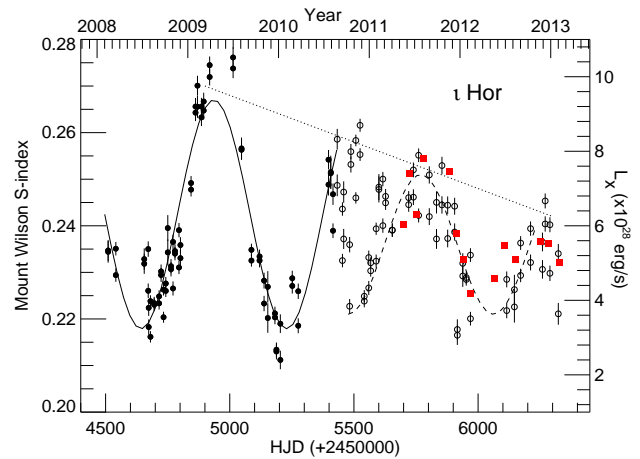}
  \caption{Time series for the Ca\,{\sc ii}\,HK $S$-index of
    \ihor. The solid line  
    represents the cycle calculated by \cite{met10} for the data shown
    with filled  circles. The new data are shown with open plotting
    symbols. A dashed line follows the new cycle starting after
    HJD\,2455500, with the same periodicity. Red filled squares 
      represent the X-ray luminosity of the coronal observations. 
    A dotted line is displayed marking the long-term trend of the
    cycle maxima.}
  \label{fig:timeseries_ca}
\end{figure}
%

\section{Results}\label{sect:results}

Figure~\ref{fig:timeseries_ca} shows the 
full \ion{Ca}{ii}\,HK time series of \ihor\, from 2008, 
the beginning of the optical spectroscopic monitoring, 
until the time of writing. 
The data collected between 2008 and the middle of 2010 are represented
in Fig.~\ref{fig:timeseries_ca}, with the 1.6~yr cycle derived for these data
by \citet{met10} overplotted. After this time span,
the clearly periodic behavior of the $S$-index was replaced by erratic
\ion{Ca}{\sc ii}\,HK variations. A few months later,
the cyclic variability seems to have returned with a lower amplitude. We
plot in Fig.~\ref{fig:timeseries_ca} the new apparent cycle, with the
same period as before but shifted in phase and with smaller
amplitude. The new cycle seems to stop near the end of
2012, when the activity decreases again instead of the
expected increase. The successive maxima, connected in
Fig.~\ref{fig:timeseries_ca}, describe a modulation by a longer term
trend, possibly a second cycle. 

\begin{figure}[t]
   \centering
   \vspace{0.5cm}
   \includegraphics[angle=90,width=0.45\textwidth]{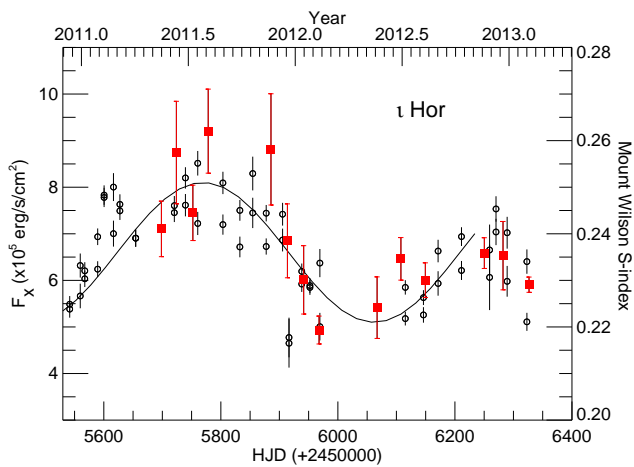}
   \caption{Time series of coronal surface flux (filled squares) and
     chromospheric S-index (open dots) for \ihor. The solid line
     indicates the cycle calculated in \citet{met10}, shifted by
     395~d. Error bars 
     of coronal X-ray surface flux are based on flux variations within each
     snapshot.}\label{fig:cycle_ca_x}
    \end{figure}

The {\em XMM-Newton} data cover the time range of the
second calcium cycle from mid-2011 until the present (February 2013). 
The mean X-ray luminosities from all snapshots are displayed together
with the \ion{Ca}{ii}\,HK 
$S$-index in Fig.~\ref{fig:timeseries_ca}, and the surface X-ray flux
is shown in Fig.~\ref{fig:cycle_ca_x}. The emission measure and
average temperature reflect the same pattern. 
The X-ray coronal emission mimics the \ion{Ca}{ii} chromospheric cycle, 
and it decreases at the end of the interval, following the
chromospheric behavior. 
The error bars for the X-ray measurements in Fig.~\ref{fig:cycle_ca_x}
represent the variations observed within each 5--11\,ks
snapshot\footnote{Light curves of each observation were generated
  using 1000~s bins (Sanz-Forcada et al., in prep.). The maximum and
  minimum of the values within each 
observation are used to establish the ``error bars'' around a central
point.}. No
strong flares are seen in these light curves, but small-scale X-ray
variability is present (Sanz-Forcada et al., in prep.). This 
short-term variability decreases at times of lower X-ray luminosity,
i.e. near the cycle minimum.
The cyclic behavior is conditioned by the longer term
trend, which hampers the calculation of the X-ray cycle length. 
The coronal cycle
does not show any essential differences with respect to its
chromospheric counterpart.

The characteristic parameters of the chromospheric (\ion{Ca}{ii}\,HK) and the
coronal (X-ray) cycle of \ihor\, observed since HJD~2455550 (end of
2010) are indicated below. The coronal activity index has an average value of
$\log L_{\rm X}/L_{\rm bol}=-5.0$ \citep[$\log{L_{\rm bol}}\,{\rm
    [erg/s]} = 33.8$,][]{Sanz11}, spanning 0.27~dex in amplitude. In
the chromosphere we observe $\log R'_{\rm HK}=-4.6$ as an average,
with 0.2~dex spanned since the end of 2010.
The parameter $R'_{\rm HK}$ is obtained 
from the $S$-index using the calibration presented 
by \cite{noyes84b} and includes subtraction of the photospheric contribution. 
%
%

\section{Discussion and conclusions}\label{sect:disc_and_concl}

Iota Hor is the as yet youngest and most active star on which both a
chromospheric and a coronal 
activity cycle have been detected so far.
Given the F8\,V spectral type and $\sim 600$\,Myr age of \ihor, 
these observations might represent a paradigm of the first
activity cycles in the life of a solar-like star.

We find that both diagnostics, X-rays and \ion{Ca}{ii}\,HK,  
reveal evidence for changes in the activity cycle of \ihor\, on a timescale
comparable to the cycle duration. The variability observed in
Fig.~\ref{fig:timeseries_ca}, and described in Sect.~3, draws a
scenario with a short (1.6~yr) cycle that shows some irregularities. 
A possible explanation is a second, longer cycle superposed 
onto the $1.6$\,yr variation and modulating the shorter cycle. 
Extending the time baseline of monitoring observations to a few
more years will shed light on this interpretation.
The presence of more than one activity cycle 
persisting on the same star at a given time has been noted in \ion{Ca}{ii}\,HK data
for other active stars \citep[e.g.,][]{baliunas95b}, while 
changes in the periodicity of the cycles have been observed in
\ion{Ca}{ii}\,HK and photometric data of a few stars, including the
Sun \citep{ola09}.

Altogether, \ihor\ has now been monitored for approximately three
cycle lengths, and there have been two episodes where the periodic
behavior was disrupted. Since some time after the disruption, 
the cycles resumes
with the same period, it seems that the dynamo process itself is
disturbed by some unknown mechanism. Interaction with a binary
companion would be a possible cause, but \ihor\ can be considered to be
a single star given its extensive spectroscopic monitoring that has
led to the discovery of its planet. 
We have noticed that the anomaly
in the \ihor\, activity cycle in 2010 occurs at the orbital
phase\footnote{$\phi$=0 is defined
  by \citet{Kuerster2000} as the time of maximum radial velocity of
  the planet as seen from Earth.} $\phi\sim0$ of the planet \ihor~b. 
This coincidence happens again
shortly after the disruption of the cycle in 2012. Although some
influence of the positions of the solar system giant planets in the
activity cycle of 
the Sun has been proposed recently \citep{abr12}, it seems unlikely to
us that the planet (at 0.9~a.u.) has any influence on the stellar cycle of
\ihor. The orbital parameters of \citet{nae01} do not show any coincidence. 
An alternative explanation would relate the observed behavior with the
geometry of the star. In the Sun, the two hemispheres can drive cycles
that are out of phase with each other \citep[e.g.][]{ribes93}, which
can lead to phenomena like the recent extended minimum. In \ihor,
with an inclination 
$i\sim60\degr$, we are primarily seeing one magnetic pole at a time, so
it appears that one of the magnetic poles may be driving a coherent
cycle, while the other is doing something more
chaotic. Because we can observe some fraction of the less visible
hemisphere, we see more than 50\% of the true cycle period as the
coherent pole drives a cycle first in the more visible hemisphere and
then in the less visible hemisphere. Eventually the activity belt
migrates out of view in the less visible hemisphere, and we are left
with only the chaotic hemisphere.

According to the relations between X-ray luminosity and rotation given by 
\cite{Pizzolato03.1} our mean value ($L_{\rm X} = 6 \times
10^{28}$\,erg/s) corresponds  
to a rotation period of $\sim9$\,d. This is in fair agreement with the
period derived 
from \ion{Ca}{ii}\,HK measurements of $\sim8$\,d \citep{met10, boi11}.
The ratio between rotation and activity cycle period has gained
attention in the  
literature because it constrains the strength of the
$\alpha$-effect in kinematic dynamo theory. 
A positive correlation between observed values for $P_{\rm rot}$ and 
$P_{\rm cyc}$ derived 
for a subsample from the Mt.Wilson project with reliable cycles 
has been established by various authors 
\citep[e.g.,][]{Brandenburg98.1, Saar99.1, BoehmVitense07}. 
These studies identified
different branches for active and inactive stars that are separated by 
$\log{R'_{\rm HK}} = -4.75$. 
A large fraction of the active group, represented by faster rotators, displays
two activity cycles. For these stars the shorter cycle lies on the extension
of the inactive branch to short rotation periods. 
This relation has been reproduced by \cite{Lorente05} 
with a local solution for an interface dynamo. 
\ihor\, fits into this picture well. Its mean chromospheric activity
($\log{\langle R'_{\rm HK} \rangle} = -4.6$) puts it in the active group. 
The measured values for its rotation ($P_{\rm rot} \sim 8$\,d) and 
cycle ($P_{\rm cyc} = 1.6$\,yr) periods are in 
excellent agreement with the linear regression derived by \cite{Lorente05}
for which \ihor\, to date defines the low-period end. 
Moreover, we found indications of a longer, as yet
unconstrained  
cycle in \ihor\, in the form of a decreasing amplitude of its
$1.6$\,yr signal.  

\begin{figure}[t]
  \centering
  \vspace{0.5cm}
  \includegraphics[angle=0,width=0.5\textwidth]{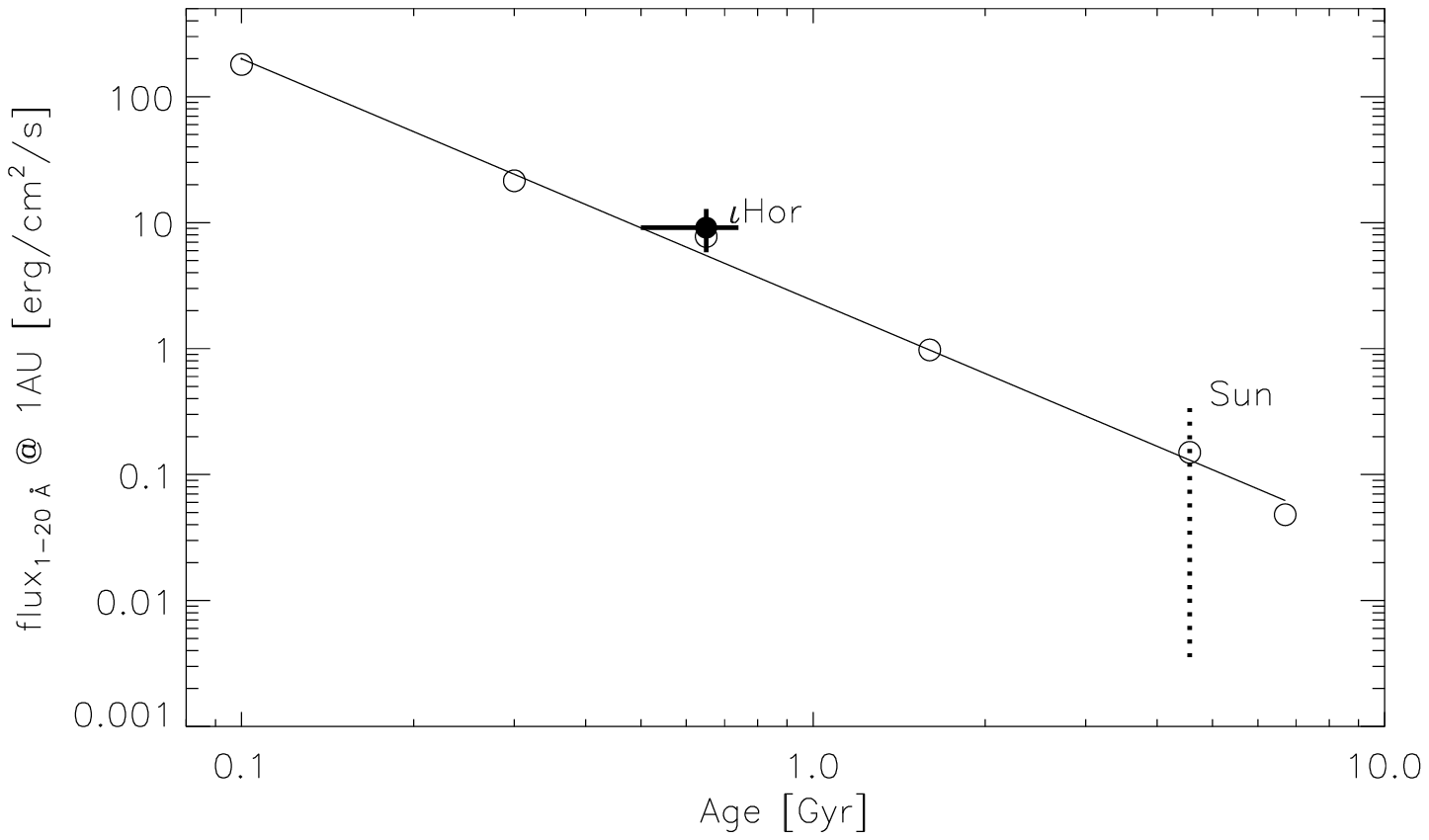}
  \caption{X-ray flux in $1-20$\,\AA~ versus stellar age for solar analogs.
The six stars studied by \protect\cite{rib05}
(including the Sun) are 
shown as open circles and the corresponding power law fit as a solid line. 
\ihor\, is displayed with a filled circle.
Vertical lines in \ihor\ and the Sun account for their activity cycle
as measured in the ASCA $0.5$--$10$\,keV band \citep{orl01}. All fluxes have been 
normalized to a distance of $1$\,AU from the star and a stellar radius of
$1\,R_\odot$. 
}
  \label{fig:fx_age}
\end{figure}

Coronal activity cycles in planet-host stars such as \ihor, 
are of particular interest given the disputed evidence of a connection
between solar cycle variations and Earth's climate \citep{fri91}. 
From an astrobiological point of view, the case of \ihor\ is
interesting because it represents a young solar-like star at an age of
$\sim$600~Myr, coincident with the time at which life started to
develop on Earth. 
The high-energy radiation from the Sun may have
played an important role during initial steps in the evolution of life
\citep[e.g.][and references therein]{cno07}. 
Winds and high-energy radiation of the young Sun 
have also been held responsible for processes on other solar system planets, 
e.g. the erosion of Mercury's early atmosphere \citep{Guinan04.0}.
Similarly the amount of UV and XUV radiation and its variation in time 
can be considered crucial for the evolution of the atmospheres and for 
the development of life on extrasolar planets \citep{Sanz11}.
\cite{rib05} have compared the XUV fluxes of six solar analogs,
including the Sun itself,
spanning an age range from $100$\,Myr to $6.7$\,Gyr. They
observed a power-law decrease of the high-energy emission for all 
wavelength ranges between $1$\AA\, and $\sim1200$\,\AA. 
Figure~\ref{fig:fx_age} shows
\ihor, together with the data from 
\cite{rib05}, for the X-ray band ($1$--$20$\,\AA~ range).
The position of \ihor\, agrees well with this relation.

Activity cycles add a shorter term modulation to the amount of high-energy
radiation arriving in the planet atmosphere. 
An active star like
\ihor\ is expected to have a large area of its surface covered with
active regions even at the minimum of its cycle. This might
explain why a coronal cycle only results in an increase in the X-ray
emission by a factor of two. 
It is presently unclear if the amplitude of these modulations 
depends on stellar age, since only two solar-like stars with known
age have measured X-ray activity cycles.  
For \ihor, this variation may be minor 
compared to large flares and star-to-star variations.
In contrast, almost two orders of magnitude are spanned by the solar 
coronal cycle (see Fig.~\ref{fig:fx_age}).
Finally, the \ihor\ system may in the future allow us to
track the effects of the stellar radiation on a giant planet at the
Sun-Earth distance.

\begin{acknowledgements}
We acknowledge Norbert Schartel for the last observation, granted as
XMM-Newton Director Discretionary Time (DDT). JSF acknowledge support from the
  Spanish MICINN through grant AYA2008-02038 and AYA2011-30147-C03-03.
The Southern HK project was supported under the NOAO long-term program
2011B-0001 with additional time from SMARTS partner institutions.
BS acknowledges support by the Faculty of the European Space Astronomy
Center (ESAC). We also thank B. Montesinos and Billy T. Orlando for useful
conversations on the topic.
\end{acknowledgements}

\Online

\end{document}